\documentclass[journal]{IEEEtran}
\usepackage{amsfonts,cite,latexsym}

\newcommand{\escape}
        {\ensuremath{\bot}}

\newcommand{\cost}[1]
	{\ensuremath{\mathrm{cost} (#1)}}

\newtheorem{theorem}{Theorem}
\newtheorem{lemma}[theorem]{Lemma}

\begin{document}

\title{Dynamic Shannon Coding}
\author{Travis~Gagie,~\IEEEmembership{Student Member,~IEEE}
\thanks{Travis Gagie is with the Department of Computer Science, 
University of Toronto (email: {\tt travis@cs.toronto.edu}).}}
\maketitle

\begin{abstract}
We present a new algorithm for dynamic prefix-free coding, based on
Shannon coding.  We give a simple analysis and prove a better upper bound
on the length of the encoding produced than the corresponding bound for
dynamic Huffman coding.  We show how our algorithm can be modified for
efficient length-restricted coding, alphabetic coding and coding with
unequal letter costs.
\end{abstract}


\section{Introduction}

Prefix-free coding is a well-studied problem in data compression and
combinatorial optimization.  For this problem, we are given a string \(S =
s_1 \cdots s_m\) drawn from an alphabet of size $n$ and must encode each
character by a self-delimiting binary codeword.  Our goal is to minimize
the length of the entire encoding of $S$.  For static prefix-free coding,
we are given all of $S$ before we start encoding and must encode every
occurrence of the same character by the same codeword.  The assignment of
codewords to characters is recorded as a preface to the encoding.  For
dynamic prefix-free coding, we are given $S$ character by character and
must encode each character before receiving the next one.  We can use a
different codeword for different occurrences of the same character, we do
not need a preface to the encoding and the assignment of codewords to
characters cannot depend on the suffix of $S$ not yet encoded.

The best-known algorithms for static coding are by
\mbox{Shannon~\cite{Sha48}} and Huffman~\cite{Huf52}.  Shannon's algorithm
uses at most \((H + 1) m + O (n \log n)\) bits to encode $S$, where
\[H = \sum_{a \in S}
	\frac{\#_a (S)}{m} \log \left( \frac{m}{\#_a (S)} \right)\]
is the empirical entropy of $S$ and \(\#_a (S)\) is the number of
occurrences of the character $a$ in $S$.  By $\log$ we mean $\log_2$.  
Shannon proved a lower bound of \(H m\) bits for all coding algorithms, 
whether or not they are prefix-free.  Huffman's algorithm produces an 
encoding that, excluding the preface, has minimum length.  The total 
length is \((H + r) m + O (n \log n)\) bits, where \(0 \leq r < 1\) is a 
function of the character frequencies in $S$~\cite{DD96}.

Both algorithms assign codewords to characters by constructing a
\emph{code-tree}, that is, a binary tree whose left and right edges are
labelled by 0's and 1's, respectively, and whose leaves are labelled by
the distinct characters in $S$.  The codeword assigned to a character $a$
in $S$ is the sequence of edge labels on the path from the root to the
leaf labelled $a$.  Shannon's algorithm builds a code-tree in which, for
\(a \in S\), the leaf labelled $a$ is of depth at most \(\lceil \log (m /
\#_a (S)) \rceil\).  Huffman's algorithm builds a Huffman tree for the
frequencies of the characters in $S$.  A \emph{Huffman tree} for a
sequence of weights \(w_1, \ldots, w_n\) is a binary tree whose leaves, in
some order, have weights \(w_1, \ldots, w_n\) and that, among all such
trees, minimizes the weighted external path length.  To build a Huffman
tree for \(w_1, \ldots, w_n\), we start with $n$ trees, each consisting of
just a root.  At each step, we make the two roots with smallest weights,
$w_i$ and $w_j$, into the children of a new root with weight \(w_i +
w_j\).

A \emph{minimax tree} for a sequence of weights \(w_1, \ldots, w_n\) is a
binary tree whose leaves, in some order, have weights \(w_1, \ldots, w_n\)
and that, among all such trees, minimizes the maximum sum of any leaf's
weight and depth.  Golumbic~\cite{Gol76} gave an algorithm, similar to
Huffman's, for constructing a minimax tree.  The difference is that, when
we make the two roots with smallest weights, $w_i$ and $w_j$, into the
children of a new root, that new root has weight \(\max (w_i, w_j) + 1\)
instead of \(w_i + w_j\).  Notice that, if there exists a binary tree
whose leaves, in some order, have depths \(d_1, \ldots, d_n\), then a
minimax tree $T$ for \(- d_1, \ldots, - d_n\) is such a tree and, more
generally, the depth of each node in $T$ is bounded above by the negative
of its weight.  So we can construct a code-tree for Shannon's algorithm by
running Golumbic's algorithm, starting with roots labelled by the distinct
characters in $S$, with the root labelled $a$ having weight \(- \lceil
\log (m / \#_a (S)) \rceil\).

Both Shannon's algorithm and Huffman's algorithm have three phases: a
first pass over $S$ to count the occurrences of each distinct character,
an assignment of codewords to the distinct characters in $S$ (recorded as
a preface to the encoding) and a second pass over $S$ to encode each
character in $S$ using the assigned codeword.  The first phase takes \(O
(m)\) time, the second \(O (n \log n)\) time and the third \(O ((H + 1)
m)\) time.

For any static algorithm $\mathcal{A}$, there is a simple dynamic
algorithm that recomputes the code-tree from scratch after reading each
character.  Specifically, for \(i = 1 \ldots m\):
\begin{enumerate}
\item We keep a running count of the number of occurrences of each     
distinct character in the current prefix \(s_1 \cdots s_{i - 1}\) of $S$.
\item We compute the assignment of codewords to characters that would
result from applying $\mathcal{A}$ to \(\escape s_1 \cdots s_{i - 1}\),
where \escape\ is a special character not in the alphabet.
\item If $s_i$ occurs in \(s_1 \cdots s_{i - 1}\), then we encode $s_i$ as
the codeword $c_i$ assigned to that character.
\item If $s_i$ does not occur in \(s_1 \cdots s_{i - 1}\), then we encode
$s_i$ as the concatenation $c_i$ of the codeword assigned to \escape\ 
and the binary representation of $s_i$'s index in the alphabet.
\end{enumerate}
We can later decode character by character.  That is, we can recover \(s_1
\cdots s_i\) as soon as we have received \(c_1 \cdots c_i\).  To see why,
assume that we have recovered \(s_1 \cdots s_{i - 1}\).  Then we can
compute the assignment of codewords to characters that $\mathcal{A}$ used
to encode $s_i$.  Since $\mathcal{A}$ is prefix-free, $c_i$ is the only
codeword in this assignment that is a prefix of \(c_i \cdots c_m\).  
Thus, we can recover $s_i$ as soon as $c_i$ has been received.  This takes 
the same amount of time as encoding $s_i$.

Faller~\cite{Fal73} and Gallager~\cite{Gal78} independently gave a dynamic
coding algorithm based on Huffman's algorithm.  Their algorithm is similar
to, but much faster than, the simple dynamic algorithm obtained by
adapting Huffman's algorithm as described above.  After encoding each
character of $S$, their algorithm merely updates the Huffman tree rather
than rebuilding it from scratch.  Knuth~\cite{Knu85} implemented their
algorithm so that it uses time proportional to the length of the encoding
produced.  For this reason, it is sometimes known as Faller-Gallager-Knuth
coding; however, it is most often called \emph{dynamic \mbox{Huffman}
coding}.  Milidi\'u, Laber, and Pessoa~\cite{MLP99} showed that this
version of dynamic Huffman coding uses fewer than \(2 m\) more bits to
encode $S$ than Huffman's algorithm.  Vitter~\cite{Vit87} gave an improved
version that he showed uses fewer than $m$ more bits than Huffman's
algorithm.  These results imply Knuth's and Vitter's versions use at most
\((H + 2 + r) m + O (n \log n)\) and \((H + 1 + r) m + O (n \log n)\) bits
to encode $S$, but it is not clear whether these bounds are tight.  Both
algorithms use \(O ((H + 1) m)\) time.

In this paper, we present a new dynamic algorithm, \emph{dynamic Shannon
coding}.  In Section~\ref{simple_section}, we show that the simple dynamic
algorithm obtained by adapting Shannon's algorithm as described above,
uses at most \((H + 1)  m + O (n \log m)\) bits and \(O (m n \log n)\)
time to encode $S$.  Section~\ref{efficient_section} contains our main
result, an improved version of dynamic Shannon coding that uses at most
\((H + 1) m + O (n \log m)\) bits to encode $S$ and only \(O ((H + 1) m +
n \log^2 m)\) time.  The relationship between Shannon's algorithm and this 
algorithm is similar to that between Huffman's algorithm and dynamic 
Huffman coding, but our algorithm is much simpler to analyze than dynamic 
Huffman coding.

In Section~\ref{variations_section}, we show that dynamic Shannon coding
can be applied to three related problems.  We give algorithms for dynamic
length-restricted coding, dynamic alphabetic coding and dynamic coding
with unequal letter costs.  Our algorithms have better bounds on the
length of the encoding produced than were previously known.  For
length-restricted coding, no codeword can exceed a given length.  For
alphabetic coding, the lexicographic order of the codewords must be the
same as that of the characters.

Throughout, we make the common simplifying assumption that \(m \geq n\).  
Our model of computation is the unit-cost word RAM with \(\Omega (\log
m)\)-bit words.  In this model, ignoring space required for the input and
output, all the algorithms mentioned in this paper use \(O (|\{a: a \in
S\}|)\) words, that is, space proportional to the number of distinct
characters in $S$.

\section{Analysis of Simple Dynamic Shannon Coding}
\label{simple_section}

In this section, we analyze the simple dynamic algorithm obtained by
repeating Shannon's algorithm after each character of the string \(\escape
s_1 \cdots s_m\), as described in the introduction.  Since the second
phase of Shannon's algorithm, assigning codewords to characters, takes \(O
(n \log n)\) time, this simple algorithm uses \(O (m n \log n)\) time to
encode $S$.  The rest of this section shows this algorithm uses at most
\((H + 1) m + O (n \log m)\) bits to encode $S$.

For \(1 \leq i \leq m\) and each distinct character $a$ that occurs in
\(\escape s_1 \cdots s_{i - 1}\), Shannon's algorithm on \(\escape s_1
\cdots s_{i - 1}\) assigns to $a$ a codeword of length at most \(\lceil
\log (i / \#_a (\escape s_1 \cdots s_{i - 1})) \rceil\).  This fact is
key to our analysis.

Let $R$ be the set of indices $i$ such that $s_i$ is a repetition of a
character in \(s_1 \cdots s_{i - 1}\).  That is, \(R = \{i : 1 \leq i \leq
m,\ s_i \in \{s_1, \ldots, s_{i - 1}\}\}\).  Our analysis depends on the
following technical lemma.

\begin{lemma}
\label{analysis_lemma_1}
\[\sum_{i \in R}
	\log \left( \frac{i}{\#_{s_i} (s_1 \cdots s_{i - 1})} \right)
\leq H m + O (n \log m)\ .\]
\end{lemma}

\begin{proof}
Let
\[L
= \sum_{i \in R}
	\log \left( \frac{i}{\#_{s_i} (s_1 \cdots s_{i - 1})} \right)\ .\]
Notice that \(\sum_{i \in R} \log i < \sum_{i = 1}^m \log i = \log
(m!)\). Also, for \(i \in R\), if $s_i$ is the $j$th occurrence of $a$ in
$S$, for some \(j \geq 2\), then \(\log \#_{s_i} (s_1 \cdots s_{i - 1}) 
= \log (j - 1)\).  Thus,
\begin{eqnarray*}
L
& = & \sum_{i \in R} \log i - \sum_{i \in R}
	\log \#_{s_i} (s_1 \cdots s_{i - 1}) \\
& < & \log (m!) -
	\sum_{a \in S} \sum_{j = 2}^{\#_a (S)} \log (j - 1) \\
& = & \log (m!) -
	\sum_{a \in S} \log (\#_a (S)!) +
	\sum_{a \in S} \log \#_a (S)\ .
\end{eqnarray*}
There are at most $n$ distinct characters in $S$ and each occurs at most 
$m$ times, so \(\sum_{a \in S} \log \#_a (S) \in O (n \log m)\).  By 
Stirling's Formula,
\[x \log x - x \ln 2
< \log (x!)
\leq x \log x - x \ln 2 + O (\log x)\ .\]
Thus,
\begin{eqnarray*}
L
& < & m \log m - m \ln 2 - \nonumber \\
&&	\sum_{a \in S} \left( \rule{0ex}{2.5ex}
	\#_a (S) \log \#_a (S) - \#_a (S) \ln 2 \right) +
	O (n \log m)\ .
\end{eqnarray*}
Since \(\sum_{a \in S} \#_a (S) = m\),
\[L
< \sum_{a \in S} \#_a (S) \log \left( \frac{m}{\#_a (S)} \right) +
	O (n \log m)\ .\]
By definition, this is \(H m + O (n \log m)\).
\end{proof}

As an aside, we note \(\sum_{a \in S} \log \#_a (S) \in o ((H + 1) m)\); 
to see why, compare corresponding terms in \(\sum_{a \in S} \log \#a (S)\) 
and the expansion
\[(H + 1) m
= \sum_{a \in S} \#_a (S)
	\left( \log \left( \frac{m}{\#_a (S)} \right) + 1 \right)\ .\]

Using Lemma~\ref{analysis_lemma_1}, it is easy to bound the number of bits
that simple dynamic Shannon coding uses to encode $S$.

\begin{theorem}
Simple dynamic Shannon coding uses at most \((H + 1) m + O (n \log m)\) 
bits to encode $S$.
\end{theorem}

\begin{proof}
If $s_i$ is the first occurrence of that character in $S$ (i.e., \(i \in
\{1, \ldots, m\} - R\)), then the algorithm encodes $s_i$ as the codeword
for \escape, which is at most \(\lceil \log m \rceil\) bits, followed by
the binary representation of $s_i$'s index in the alphabet, which is
\(\lceil \log n \rceil\) bits.  Since there are at most $n$ such 
characters, the algorithm encodes them all using \(O (n \log m)\) bits.

Now, consider the remaining characters in $S$, that is, those 
characters whose indices are in $R$.  In total, the algorithm encodes 
these using at most
\begin{eqnarray*}
\lefteqn{\sum_{i \in R} \left\lceil
	\log \left( \frac{i}{\#_{s_i} (\escape s_1 \cdots s_{i - 1})} \right)
	\right\rceil} \\
& < & m + \sum_{i \in R}
	\log \left( \frac{i}{\#_{s_i} (s_1 \cdots s_{i - 1})} \right)
\end{eqnarray*}
bits.  By Lemma~\ref{analysis_lemma_1}, this is at most \((H + 1) m + O (n 
\log m)\).

Therefore, in total, this algorithm uses at most \((H + 1) m + O (n \log 
m)\) bits to encode $S$.
\end{proof}

\section{Dynamic Shannon Coding}
\label{efficient_section}

This section explains how to improve simple dynamic Shannon coding so that
it uses at most \((H + 1) m + O (n \log m)\) bits and \(O ((H + 1) m + n
\log^2 m)\)  time to encode the string \(S = s_1 \cdots s_m\).  The main
ideas for this algorithm are using a dynamic minimax tree to store the
code-tree, introducing ``slack'' in the weights and using background
processing to keep the weights updated.

Gagie~\cite{Gag03} showed that Faller's, Gallager's and Knuth's
techniques for making Huffman trees dynamic can be used to make minimax
trees dynamic.  A \emph{dynamic minimax tree} $T$ supports the following 
operations:
\begin{itemize}
\item given a pointer to a node $v$, return $v$'s parent, left child, and
right child (if they exist);
\item given a pointer to a leaf $v$, return $v$'s weight;
\item given a pointer to a leaf $v$, increment $v$'s weight;
\item given a pointer to a leaf $v$, decrement $v$'s weight;
\item and, given a pointer to a leaf $v$, insert a new leaf with the same
	weight as $v$.
\end{itemize}
In Gagie's implementation, if the depth of each node is bounded above by
the negative of its weight, then each operation on a leaf with weight 
$-d_i$ takes \(O (d_i)\) time.  Next, we will show how to use this data 
structure for fast dynamic Shannon coding.

We maintain the invariant that, after we encode \(s_1 \cdots s_{i - 1}\),
$T$ has one leaf labelled $a$ for each distinct character $a$ in \(\escape
s_1 \cdots s_{i - 1}\) and this leaf has weight between \(- \lceil \log
((i + n) / \#_a (\escape s_1 \cdots s_{i - 1})) \rceil\) and \(- \lceil
\log (\max (i, n) / \#_a (\escape s_1 \cdots s_{i - 1})) \rceil\).  
Notice that applying Shannon's algorithm to \(\escape s_1 \cdots s_{i -
1}\) results in a code-tree in which, for \(a \in \escape s_1 \cdots s_{i
-1}\), the leaf labelled $a$ is of depth at most \(\lceil \log (i / \#_a 
(\escape s_1 \cdots s_{i - 1})) \rceil\).  It follows that the depth of 
each node in $T$ is bounded above by the negative of its weight.

Notice that, instead of having just $i$ in the numerator, as we would for
simple dynamic Shannon coding, we have at most \(i + n\).  Thus, this
algorithm may assign slightly longer codewords to some characters.  We
allow this ``slack'' so that, after we encode each character, we only need
to update the weights of at most two leaves. In the analysis, we will show
that the extra $n$ only affects low-order terms in the bound on the length
of the encoding.

After we encode $s_i$, we ensure that $T$ contains one leaf labelled $s_i$
and this leaf has weight \(- \lceil \log ((i + 1 + n) / \#_{s_i} (\escape
s_1 \cdots s_i)) \rceil\).  First, if $s_i$ is the first occurrence of
that distinct character in $S$ (i.e., \(i \in \{1, \ldots, m\} - R\)),
then we insert a new leaf labelled $s_i$ into $T$ with the same weight as
the leaf labelled \escape. Next, we update the weight of the leaf labelled
$s_i$.  We consider this processing to be in the foreground.

In the background, we use a queue to cycle through the distinct characters
that have occurred in the current prefix.  For each character that we
encode in the foreground, we process one character in the background. When
we dequeue a character $a$, if we have encoded precisely \(s_1 \cdots
s_i\), then we update the weight of the leaf labelled $a$ to be \(- \lceil
\log ((i + 1 + n) / \#_a (\escape s_1 \cdots s_i)) \rceil\), unless it has
this weight already.  Since there are always at most \(n + 1\)  distinct
characters in the current prefix (\escape\ and the $n$ characters in the
alphabet), this maintains the following invariant:  For \(1 \leq i \leq
m\) and \(a \in \escape s_1 \cdots s_{i - 1}\), immediately after we
encode \(s_1 \cdots s_{i - 1}\), the leaf labelled $a$ has weight between
\(- \lceil \log ((i + n) / \#_a (\escape s_1 \cdots s_{i - 1})) \rceil\)
and \(- \lceil \log (\max (i, n) / \#_a (\escape s_1 \cdots s_{i - 1}))
\rceil\).  Notice that \(\max (i, n) < i + n \leq 2 \max (i, n)\) and
\(\#_a (s_1 \cdots s_{i - 1}) \leq \#_a (\escape s_1 \cdots s_i) \leq 2
\#_a (s_1 \cdots s_{i - 1}) + 1\).  Also, if $s_i$ is the first occurrence
of that distinct character in $S$, then \(\#_{s_i} (\escape s_1 \cdots
s_i) = \#_{\escape} (\escape s_1 \cdots s_{i - 1})\).  It follows that,
whenever we update a weight, we use at most one increment or decrement.

Our analysis of this algorithm is similar to that in
Section~\ref{simple_section}, with two differences.  First, we show that
weakening the bound on codeword lengths does not significantly affect the
bound on the length of the encoding.  Second, we show that our algorithm
only takes \(O ((H + 1) m + n \log^2 m)\) time.  Our analysis depends on
the following technical lemma.

\begin{lemma}
\label{analysis_lemma_2}
Suppose \(I \subseteq {\mathbb Z^+}\) and \(|I| \geq n\).  Then
\[\sum_{i \in I} \log \left( \frac{i + n}{x_i} \right)
\leq \sum_{i \in I} \log \left( \frac{i}{x_i} \right) +
	n \log (\max I + n)\ .\]
\end{lemma}

\begin{proof}
Let
\[L
= \sum_{i \in I} \log \left( \frac{i + n}{x_i} \right)
= \sum_{i \in I} \log \left( \frac{i}{x_i} \right) +
	\sum_{i \in I} \log \left( \frac{i + n}{i} \right)\ .\]
Let \(i_1, \ldots, i_{|I|}\) be the elements of $I$, with \(0 < i_1 < 
\cdots < i_{|I|}\).  Then \(i_j + n \leq i_{j + n}\), so
\begin{eqnarray*}
\lefteqn{\sum_{i \in I} \log \left( \frac{i + n}{i} \right)} \\
& = & \log \left( \frac
	{\left( \prod_{j = 1}^{|I| - n} (i_j + n) \right)
	\left( \prod_{j = |I| - n + 1}^{|I|} (i_j + n) \right)}
	{\left( \prod_{j = 1}^n i_j \right)
	\left( \prod_{j = n + 1}^{|I|} i_j \right)}
	\right) \\
& \leq & \log \left( \frac
	{\left( \prod_{j = 1}^{|I| - n} i_{j + n} \right)
	(\max I + n)^n}
	{1 \cdot \prod_{j = 1}^{|I| - n} i_{j + n}}
	\right) \\
& = & n \log (\max I + n)\ .
\end{eqnarray*}
Therefore,
\[L
\leq \sum_{i \in I} \log \left( \frac{i}{x_i} \right) +
	n \log (\max I + n)\ . \]
\end{proof}

Using Lemmas~\ref{analysis_lemma_1} and~\ref{analysis_lemma_2}, it is easy
to bound the number of bits and the time dynamic Shannon coding uses to
encode $S$, as follows.

\begin{theorem}  
\label{efficient_theorem}
Dynamic Shannon coding uses at most \((H + 1) m + O (n \log m)\) bits and 
\(O ((H + 1) m + n \log^2 m)\) time.   
\end{theorem}
        
\begin{proof}
First, we consider the length of the encoding produced.  Notice that the
algorithm encodes $S$ using at most
\begin{eqnarray*}
&& \sum_{i \in R} \left\lceil \log
	\left( \frac{i + n}{\#_{s_i} (\escape s_1 \cdots s_{i - 1})} \right)
	\right\rceil + O (n \log m) \\
& \leq & m + \sum_{i \in R} \log
	\left( \frac{i + n}{\#_{s_i} (s_1 \cdots s_{i - 1})} \right) +
	O (n \log m)
\end{eqnarray*}
bits.  By Lemmas~\ref{analysis_lemma_1} and~\ref{analysis_lemma_2}, this
is at most \((H + 1) m + O (n \log m)\).

Now, we consider how long this algorithm takes.  We will prove separate
bounds on the processing done in the foreground and in the background.

If $s_i$ is the first occurrence of that character in $S$ (i.e., \(i \in
\{1, \ldots, m\} - R\)), then we perform three operations in the
foreground when we encode $s_i$: we output the codeword for \escape, which
is at most \(\lceil \log (i + n) \rceil\) bits; we output the index of
$s_i$ in the alphabet, which is \(\lceil \log n \rceil\) bits; and we
insert a new leaf labelled $s_i$ and update its weight to be \(- \lceil
\log (i + 1 + n) \rceil\).  In total, these take \(O (\log (i + n))  
\subseteq O (\log m)\) time.  Since there are at most $n$ such characters,
the algorithm encodes them all using \(O (n \log m)\) time.

For \(i \in R\), we perform at most two operations in the foreground when
we encode $s_i$: we output the codeword for $s_i$, which is of length at
most \linebreak \(\lceil \log ((i + n) / \#_{s_i} (s_1 \cdots s_{i - 1}))  
\rceil\); and, if necessary, we increment the weight of the leaf labelled
$s_i$.  In total, these take \(O \left( \log ((i + n) / \#_{s_i} (s_1
\cdots s_{i - 1})) \right)\) time.

For \(1 \leq i \leq m\), we perform at most two operations in the
background when we encode $s_i$: we dequeue a character $a$; if necessary,  
decrement the weight of the leaf labelled $a$; and re-enqueue $a$.  These
take \(O (1)\) time if we do not decrement the weight of the leaf labelled
$a$ and \(O (\log m)\) time if we do.

Suppose $s_i$ is the first occurrence of that distinct character in
$S$.  Then the leaf $v$ labelled $s_i$ is inserted into $T$ with weight
\(- \lceil \log (i + n) \rceil\).  Also, $v$'s weight is never less than
\(- \lceil \log (m + 1 + n) \rceil\).  Since decrementing $v$'s weight
from $w$ to \(w - 1\) or incrementing $v$'s weight from \(w - 1\) to $w$
both take \(O (- w)\) time, we spend the same amount of time decrementing
$v$'s weight in the background as we do incrementing it in the foreground,
except possibly for the time to decrease $v$'s weight from \(- \lceil \log
(i + n) \rceil\) to \(- \lceil \log (m + 1 + n) \rceil\).  Thus, we spend
\(O (\log^2 m)\) more time decrementing $v$'s weight than we do
incrementing it.  Since there are at most $n$ distinct characters in $S$,
in total, this algorithm takes
\[\sum_{i \in R} O \left( \log \left(
	\frac{i + n}{\#_{s_i} (s_1 \cdots s_{i - 1})}
	\right) \right) +
	O (n \log^2 m)\]
time.  It follows from Lemmas~\ref{analysis_lemma_1}
and~\ref{analysis_lemma_2} that this is \(O ((H + 1) m + n \log^2 m)\).
\end{proof}

\section{Variations on Dynamic Shannon Coding}
\label{variations_section}

In this section, we show how to implement efficiently variations of
dynamic Shannon coding for dynamic length-restricted coding, dynamic
alphabetic coding and dynamic coding with unequal letter costs.  
Abrahams~\cite{Abr01} surveys static algorithms for these and similar
problems, but there has been relatively little work on dynamic algorithms
for these problems.

We use dynamic minimax trees for length-restricted dynamic Shannon coding.  
For alphabetic dynamic Shannon coding, we dynamize Melhorn's version of
Shannon's algorithm.  For dynamic Shannon coding with unequal letter
costs, we dynamize Krause's version.

\subsection{Length-Restricted Dynamic Shannon Coding}

For length-restricted coding, we are given a bound and cannot use a
codeword whose length exceeds this bound.  Length-restricted coding is
useful, for example, for ensuring that each codeword fits in one machine
word.  Liddell and Moffat~\cite{LM01} gave a length-restricted dynamic
coding algorithm that works well in practice, but it is quite complicated
and they did not prove bounds on the length of the encoding it produces.  
We show how to length-restrict dynamic Shannon coding without
significantly increasing the bound on the length of the encoding produced.

\begin{theorem}
\label{restriction_theorem}
For any fixed integer \(\ell \geq 1\), dynamic Shannon coding can be
adapted so that it uses at most \(2 \lceil \log n \rceil + \ell\)  bits
to encode the first occurrence of each distinct character in $S$, at most
\(\lceil \log n \rceil + \ell\) bits to encode each remaining character in
$S$, at most \(\left( H + 1 + \frac{1}{(2^{\ell} - 1) \ln 2} \right) m + O
(n \log m)\) bits in total, and \(O ((H + 1) m + n \log^2 m)\) time.
\end{theorem}

\begin{proof}
We modify the algorithm presented in Section~\ref{efficient_section} by
removing the leaf labelled \escape\ after all of the characters in the
alphabet have occurred in $S$, and changing how we calculate weights for 
the dynamic minimax tree.  Whenever we would use a weight of the form \(-
\lceil \log x \rceil\), we smooth it by instead using
\begin{eqnarray*}
\lefteqn{- \left\lceil \log \left(
	\frac{2^{\ell}}{(2^{\ell} - 1) / x + 1 / n}
	\right) \right\rceil} \\
& \geq & - \min \left(
	\left\lceil \log \left( \frac{2^{\ell} x}{2^{\ell} - 1} \right) 
		\right\rceil,
	\left\lceil \log n \right\rceil + \ell
	\right)\ .
\end{eqnarray*}
With these modifications, no leaf in the minimax tree is ever of depth
greater than \(\lceil \log n \rceil + \ell\).  Since
\begin{eqnarray*}
\left\lceil \log \left( \frac{2^{\ell} x}{2^{\ell} - 1} \right) \right\rceil
& < & \log x + 1 + \frac{\log \left( 1 +
	\frac{1}{2^{\ell} - 1} \right)^{2^{\ell} - 1}}{2^{\ell} - 1} \\
& < & \log x + 1 + \frac{1}{(2^{\ell} - 1) \ln 2}\ ,
\end{eqnarray*}
essentially the same analysis as for Theorem~\ref{efficient_theorem} shows 
this algorithm uses at most \(\left( H + 1 + \frac{1}{(2^{\ell} - 1) \ln 
2} \right) m + O (n \log m)\) bits in total, and \(O ((H + 1) m + n \log^2 
m)\) time.
\end{proof}

It is straightforward to prove a similar theorem in which the number of
bits used to encode $s_i$ with \(i \in R\) is bounded above by \(\lceil
\log (|\{a : a \in S\}| + 1) \rceil + \ell + 1\) instead of \(\lceil \log
n \rceil + \ell\).  That is, we can make the bound in terms of the number
of distinct characters in $S$ instead of the size of the alphabet.  To do
this, we modify the algorithm again so that it stores a counter $n_i$ of
the number of distinct characters that have occurred in the current
prefix.  Whenever we would use $n$ in a formula to calculate a weight, we
use \(2 (n_i + 1)\) instead.

\subsection{Alphabetic Dynamic Shannon Coding}
\label{alphabetic_subsection}

For alphabetic coding, the lexicographic order of the codewords must
always be the same as the lexicographic order of the characters to which
they are assigned.  Alphabetic coding is useful, for example, because we
can compare encoded strings without decoding them. Although there is an
alphabetic version of minimax trees~\cite{KK85}, it cannot be efficiently
dynamized~\cite{Gag03}.  Mehlhorn~\cite{Meh77} generalized Shannon's
algorithm to obtain an algorithm for alphabetic coding.  In this section,
we dynamize Mehlhorn's algorithm.

\begin{theorem}[Mehlhorn, 1977]
There exists an alphabetic prefix-free code such that, for each character 
$a$ in the alphabet, the codeword for $a$ is of length \hfill \linebreak 
\(\lceil \log ((m + n) / \#_a (S)) \rceil + 1\).
\end{theorem}

\begin{proof}
Let \(a_1, \ldots, a_n\) be the characters in the alphabet in 
lexicographic order.  For \(1 \leq i \leq n\), let
\[f (a_i)
= \frac{\#_{a_i} (S) + 1}{2 (m + n)} +
	\sum_{j = 1}^{i - 1} \frac{\#_{a_j} (S) + 1}{m + n}
< 1\ .\]
For \(1 \leq i \neq i' \leq n\), notice that \(|f (a_i) - f (a_{i'})| \geq
\frac{\#_{a_i} (S) + 1}{2 (m + n)}\).  Therefore, the first \(\left\lceil
\log \left(\frac{m + n}{\#_{a_i} (S) + 1} \right) \right\rceil + 1\) bits
of the binary representation of \(f (a_i)\) suffice to distinguish it.  
Let this sequence of bits be the codeword for $a_i$.
\end{proof}

Repeating Mehlhorn's algorithm after each character of $S$, as described
in the introduction, is a simple algorithm for alphabetic dynamic Shannon
coding.  Notice that we always assign a codeword to every character in the
alphabet; thus, we do not need to prepend \escape\ to the current prefix
of $S$.  This algorithm uses at most \((H + 2) m + O (n \log m)\) bits and
\(O (m n)\) time to encode $S$.

To make this algorithm more efficient, after encoding each character of
$S$, instead of computing an entire code-tree, we only compute the
codeword for the next character in $S$.  We use an augmented splay
tree~\cite{ST85} to compute the necessary partial sums.

\begin{theorem}
\label{alphabetic_theorem}
Alphabetic dynamic Shannon coding uses \((H + 2) m + O (n \log m)\) bits 
and \(O ((H + 1) m)\) time.
\end{theorem}

\begin{proof}
We keep an augmented splay tree $T$ and maintain the invariant that, after 
encoding \(s_1 \cdots s_{i - 1}\), there is a node $v_a$ in $T$ for each 
distinct character $a$ in \(s_1 \ldots, s_{i - 1}\).  The node $v_a$'s key 
is $a$; it stores $a$'s frequency in \(s_1 \cdots s_{i - 1}\) and the sum 
of the frequencies of the characters in $v_a$'s subtree in $T$.

To encode $s_i$, we use $T$ to compute the partial sum
\[\frac{\#_{s_i} (s_1 \cdots s_{i - 1})}{2} +
	\sum_{a_j < s_i} \#_{a_j} (s_1 \cdots s_{i - 1})\ ,\]
where \(a_j < s_i\) means that $a_j$ is lexicographically less than $s_i$.  
From this, we compute the codeword for $s_i$, that is, the first
\(\left\lceil \log \left( \frac{i - 1 + n}
	{\#_{s_i} (s_1 \cdots s_{i - 1}) + 1} \right) \right\rceil + 1\) 
bits of the binary representation of
\[\frac{\#_{s_i} (s_1 \cdots s_{i - 1}) + 1}{2 (i - 1 + n)} +
	\sum_{a_j < s_i} \frac{\#_{a_j} (s_1 \cdots s_{i - 1}) + 1}
	{i - 1 + n}\ .\]
If $s_i$ is the first occurrence of that character in $S$ (i.e., \(i \in 
\{1, \ldots, m\} - R\)), then we insert a node $v_{s_i}$ into $T$.  In 
both cases, we update the information stored at the ancestors of $v_{s_i}$ 
and splay $v_{s_i}$ to the root. 

Essentially the same analysis as for Theorem~\ref{efficient_theorem} shows 
this algorithm uses at most \((H + 2) m + O (n \log m)\) bits.  By the 
Static Optimality theorem~\cite{ST85}, it uses \(O ((H + 1) m)\) time.
\end{proof}

\subsection{Dynamic Shannon Coding with Unequal Letter Costs}

It may be that one code letter costs more than another.  For example,
sending a dash by telegraph takes longer than sending a dot.  
Shannon~\cite{Sha48} proved a lower bound of \(H m \ln (2) / C\) for all
algorithms, whether prefix-free or not, where the \emph{channel capacity}
$C$ is the largest real root of \(e^{- \cost{0} \cdot x} + e^{- \cost{1}
\cdot x} = 1\) and \(e \approx 2.71\) is the base of the natural
logarithm.  Krause~\cite{Kra62} generalized Shannon's algorithm for the
case with unequal positive letter costs.  In this section, we dynamize 
Krause's algorithm.

\begin{theorem}[Krause, 1962]
Suppose \cost{0}\ and \cost{1}\ are constants with \(0 < \cost{0}
\leq \cost{1}\).  Then there exists a prefix-free code such that, for each 
character $a$ in the alphabet, the codeword for $a$ has cost less than
\(\frac{\ln (m / \#_a (S))}{C} + \cost{1}\).
\end{theorem}

\begin{proof}
Let \(a_1, \ldots, a_k\) be the characters in $S$ in non-increasing order 
by frequency.  For \(1 \leq i \leq k\), let
\[f (a_i)
= \sum_{j = 1}^{i - 1} \frac{\#_{a_j} (S)}{m}
< 1\ .\]

Let \(b (a_i)\) be the following binary string, where \(x_0 = 0\) and
\(y_0 = 1\):  For \(j \geq 1\), if \(f (a_i)\) is in the first $e^{-
\cost{0} \cdot C}$ fraction of the interval \([x_{j - 1}, y_{j - 1})\),
then the $j$th bit of \(b (a_i)\) is 0 and $x_j$ and $y_j$ are such that
\([x_j, y_j)\) is the first $e^{- \cost{0} \cdot C}$ fraction of \([x_{j -
1}, y_{j - 1})\).  Otherwise, the $j$th bit of \(b (a_i)\) is 1 and $x_j$
and $y_j$ are such that \([x_j, y_j)\) is the last $e^{- \cost{1} \cdot
C}$ fraction of \([x_{j - 1}, y_{j - 1})\).  Notice that the cost to
encode the $j$th bit of \(b (a_i)\) is exactly \(\frac{\ln ((y_{j - 1} -
x_{j - 1}) / (y_j - x_j))}{C}\); it follows that the total cost to encode
the first $j$ bits of \(b (a_i)\) is \(\frac{\ln (1 / (y_j - x_j))}{C}\).

For \(1 \leq i \neq i' \leq k\), notice that \(|f (a_i) - f (a_{i'})| \geq 
\#_{a_i} (S) / m\).  Therefore, if \(y_j - x_j < \#_{a_i} (S) / m\), then 
the first $j$ bits of \(b (a_i)\) suffice to distinguish it.  So the 
shortest prefix of \(b (a_i)\) that suffices to distinguish \(b (a_i)\)
has cost less than \(\frac{\ln (e^{\cost{1} \cdot C} m / \#_a (S))}{C} = 
\frac{\ln (m / \#_a (S))}{C} + \cost{1}\).  Let this sequence of bits be 
the codeword for $a_i$.
\end{proof}

Repeating Krause's algorithm after each character of $S$, as described in
the introduction, is a simple algorithm for dynamic Shannon coding with
unequal letter costs.  This algorithm produces an encoding of $S$ with
cost at most \(\left( \frac{H \ln 2}{C} + \cost{1} \right) m + O (n \log
m)\) in \(O (m n)\) time.

As in Subsection~\ref{alphabetic_subsection}, we can make this simple
algorithm more efficient by only computing the codewords we need.  
However, instead of lexicographic order, we want to keep the characters in
non-increasing order by frequency in the current prefix.  We use a data
structure for dynamic cumulative probability tables~\cite{Mof99}, due to
Moffat.  This data structure stores a list of characters in non-increasing 
order by frequency and supports the following operations:
\begin{itemize}
\item given a character $a$, return $a$'s frequency;
\item given a character $a$, return the total frequency of all characters 
before $a$ in the list;
\item given a character $a$, increment $a$'s frequency; and,
\item given an integer $k$, return the last character $a$ in the list such 
that the total frequency of all characters before $a$ is at most $k$.
\end{itemize}
If $a$'s frequency is a $p$ fraction of the total frequency of all 
characters in the list, then an operation that is given $a$ or returns $a$ 
takes \(O (\log (1 / p))\) time.

Dynamizing Krause's algorithm using Moffat's data structure gives the
following theorem, much as dynamizing Mehlhorn's algorithm with an
augmented splay tree gave Theorem~\ref{alphabetic_theorem}.  We omit the
proof because it is very similar.

\begin{theorem}
\label{unequal_theorem}
Suppose \cost{0}\ and \cost{1}\ are constants with \(0 < \cost{0}
\leq \cost{1}\).  Then dynamic Shannon coding produces an encoding of 
$S$ with cost at most \(\left( \frac{H \ln 2}{C} + \cost{1} \right) m + O 
(n \log m)\) in \(O ((H + 1) m)\) time.
\end{theorem}

If \(\cost{0} = \cost{1} = 1\), then \(C = 1\) and
Theorem~\ref{unequal_theorem} is the same as
Theorem~\ref{efficient_theorem}.  We considered this special case first
because it is the only one in which we know how to efficiently maintain
the code-tree, which may be useful for some applications.

\section*{Acknowledgments}

Many thanks to \mbox{Julia} \mbox{Abrahams}, \mbox{Will} \mbox{Evans},
\mbox{Faith} \mbox{Fich}, \mbox{Mordecai} \mbox{Golin}, \mbox{Charlie}
\mbox{Rackoff} and \mbox{Ken} \mbox{Sevcik}.  This research was supported
by the Natural Sciences and Engineering Research Council of Canada.

\newpage

\bibliographystyle{IEEEtran.bst}
\bibliography{shannon}

\end{document}